\documentclass[11pt]{article}
\pdfoutput=1
\usepackage{graphicx,amssymb,amsfonts,amsmath,amssymb,amscd,amstext, mathrsfs}
\usepackage{color}
\makeatletter

\@addtoreset{equation}{section}
\newcommand{\be}{\begin{eqnarray}}
\newcommand{\ee}{\end{eqnarray}}
\newcommand{\ba}{\begin{array}}
\newcommand{\ea}{\end{array}}
\newcommand{\nn}{\nonumber}

\renewcommand{\(}{\Big(}
\renewcommand{\)}{\Big)}

\def \<{\langle}
\def \>{\rangle}

\makeatletter \@addtoreset{equation}{section} \makeatother

\begin{document}

\vspace{0.6cm}

\begin{center}
~\\~\\~\\
{\bf  \LARGE Complexity of Bose-Hubbard Model :\\ Quantum Phase Transition}
\vspace{2cm}

Wung-Hong Huang*\\
\vspace{0.5cm}
Department of Physics, National Cheng Kung University,\\
No.1, University Road, Tainan 701, Taiwan

\end{center}
\vspace{2cm}
\begin{center}{\bf  \large Abstract}\end{center}
The operator approach  is applied to investigate the  complexity of Bose-Hubbard model.  We present a systematic method  to expand the quantum complexity in  series of coupling constant.  We first study 2-sites system. For the ground state we can find the exact value of complexity which is the summation of all order. The complexity is divergent at critical value of  coupling constant that indicates the quantum  phase transition at this point. We then generalize the method to the N-sites closed chain and any dimensional system. The found properties are similar to those  in 2-sites system. We also  study the excited state  and present the  general formulas of  Bose-Hubbard model complexity, which shows a similar form as that in $\lambda \phi^4$ theory studied in our previous paper.  

\vspace{4cm}
\begin{flushleft}
* Retired Professor of NCKU, Taiwan. \\
* E-mail: whhwung@mail.ncku.edu.tw
\end{flushleft}
\newpage
\tableofcontents
\section{Introduction}
Complexity is a quantity expected to probe the non-local property of a quantum system besides the entangle entropy \cite{Faulkner1312, Hartman1305}.    In the context of the holography the  complexity=volume (CV) conjecture \cite{Susskind1403} and complexity=action (CA) conjecture  \cite{Brown1509,  Chapman1610} were  proposed to calculate the quantity. The complexity in there is the volume of an extremal codimension-one bulk surface and gravitational action evaluated on the Wheeler-DeWitt (WDW) patch anchored  on a time slice of the boundary respectively. 

Complexity have been studied in the quantum field theory \cite{Jefferson1707, Chapman1707, Khan1801, Hackl1803, Bhattacharyya1880, Huang1905}.  The complexity therein is defined as the number of  operations $\{\cal O^I\}$ needed to transform a reference state $|\psi_R\rangle$ to a target state $|\psi_T\rangle$.   These operators are also called as quantum gates : the more gates one needs, the more complexity the target state has.  To calculate the complexity one defines an  affine parameter ``s" associated to an unitary operator $U(s)$ and  use a set of function, $Y^I(s)$, to character the quantum circuit. The unitary operation connecting the reference state and target state is 
\be
U(s)=\vec{\cal P}\, e^{\int_0^s\,Y_I(s)\,{\cal O}_I},~~~~|\psi_R\rangle =U(0)|\psi_R\rangle,~~~~~|\psi_T\rangle =U(1)|\psi_R\rangle~~~\label{gate} \ ,
\ee
where $\vec{\cal P}$ is a time ordering along $s$. The  complexity ${\cal C}$ and circuit depth $D[U]$ (cost function) are  \cite{Jefferson1707} 
\be
{\cal C}&=&\underset{\rm \{Y^I\}}{\rm Min}\,D[U],~~~~~~D[U]=\int_0^1ds\,{\sum_I |Y^I(s)|^2} \ .~~~\label{distance}
\ee 
Above definitions were shown to  be consistent with a gravitational computation  \cite{Jefferson1707}.

The initial studies in field theory considered the Gaussian ground state wavefunctions in reference state and target state \cite{Jefferson1707,Khan1801, Hackl1803}.  The theories studied so far are  the free field theory  or exponential type  wavefunction in interacting model \cite{Bhattacharyya1880}.  In this (wavefunction) approach, since  that the excited-state wavefunction of harmonic oscillation  is not pure  exponential form  the wavefunction approach is hard to work.

 In our first  paper  \cite{Huang1905} we adopt the operator approach, in which the transformation matrix between the second quantization operators of reference state and target state is regarded as the quantum gate, to evaluate the complexity in free scalar field theory\footnote{The operator approach had also been used in \cite{Chapman1707, Hackl1803} to study the complexity of  fermion theory.}.   We examined the system in which the reference state is  two oscillators with same frequency $\omega_f$ while the  target state  is  two  oscillators with frequency $ \omega_1$ and $ \omega_2$.  The complexity in excited states is calculated and  find that  that the square of geodesic length in the general state $|{\rm N_1,N_2}\rangle$ is
\be
D_{\rm (N_1,N_2)}^2={\rm (N_1+1)}\left(\ln {\sqrt{\omega_1\over  \omega_f}}\,\right)^2 +{\rm (N_2+1)}\left(\ln {\sqrt{ \omega_2\over  \omega_f}}\,\right)^2
\ee
The results was furthermore  extended to the N couple harmonic oscillators which correspond to  the lattice version of  free scalar field.  

In our next  paper  \cite{Huang2008} we included interaction to further study the complexity using the operator approach.  We present a systematic method to evaluate the complexity of the $\lambda\phi^4$ field theory by the perturbation of small coupling constant.  We describes the lattice  scalar field as coupled oscillators.  In  two coupled oscillators, to the $\lambda^n$ order,  the square distance of excited state between target and reference state is
\be
D^{(n)2}_{(N_1,N_2)}=(N_1+1)\left(\ln \(\sqrt {R^{(n)}_1}\)\right)^2+(N_2+1)\left(\ln \(\sqrt {R^{(n)}_2}\)\right)^2
\ee
in which $R^{(n)}_1$ and $R^{(n)}_2$ are described by  simple recurrent relations
\be
R^{(n)}_1={\omega_1+{3\lambda\over 2}\,\({1+N_1R^{(n-1)}_1\over 2\omega^2_1}+{1+N_2R^{(n-1)}_2\over \omega_1\omega_2}\)\over  \omega_f+{3\lambda\over 2}\,{1+N_1\over 2\omega^2_f}},~~
R^{(n)}_2={\omega_2+{3\lambda\over 2}\({1+N_1R^{(n-1)}_1\over \omega_2\omega_1}+{1+N_2R^{(n-1)}_2\over 2\omega^2_2}\)
\over   \omega_f+{3\lambda\over 2}\,{1+N_2\over 2\omega^2_f}}  \ ~~~\label{R2}
\ee
with initial values  $R^{(0)}_i={\omega_i\over \omega_f}$. We had also generalize it to the case of N coupled oscillators which correspond to  the lattice version of  $\lambda\phi^4$ theory. 

Our method in  \cite{Huang2008} is very general and  can be applied to many-body model in  condense matter.  In this paper, along the same method,  we will  to study the complexity of Bose Hubbard model  \cite{Jaksch, Sood}.

In section II we briefly review the Bose Hubbard model.  In section III we describe the method and use it to  study the complexity of two-sites Bose-Hubbard model.  A general perturbative formula of the associated complexity is presented.  We  see that the formula of complexity of Bose-Hubbard model is very similar to that of $\lambda\phi^4$ field theory. Use it we  evaluate the   complexity of Bose-Hubbard model  and find  that the complexity is divergent at a critical value of parameter $g$, which can be identified as the quantum phase transition point.   In section IV  we extend the method to  the N-sites closed chain Bose-Hubbard model and any dimensional system.  We conclude in last section.

\section{Bose Hubbard Model}
Quantum Phase transitions is known to occur when a coupling constant $g$ in the Hamiltonian is varied across some critical value $g=g_c$ at which the ground state dependence on $g$ becomes non-analytic \cite{Sachdev, Matthias}. In this paper  we are interesting in the quantum phase transition happens in a system of Bosons in the background of a periodic potential and a two-particle repulsive interaction. The Hamiltonian is given by
\be
H = \int d^dx \left( -\frac{\hbar^2}{2m}  \psi^\dag(x) \nabla^2 \psi(x)\right)  + \frac{{\cal U}_0}{2} \int d^dx \psi^\dag(x) \psi^\dag(x) \psi(x) \psi(x)
\label{a1}
\ee
 The constant ${\cal U}_0$ is the short-ranged repulsion between the Bosons. At low energies, we can keep only the lowest vibrational state at each minima of ${\cal V}_0(x)$ and the dynamics is given by the Bose Hubbard model \cite{Jaksch}:
\be
{\cal H} &=& {J\over N}\sum_{\langle i,j \rangle} \tilde a^\dag_i\tilde a_j+ {1\over N}\sum_i \epsilon_i \tilde n_i + {U\over N}\sum_{i} \tilde n_i (\tilde n_i - 1),~~~~~~~\tilde n_i=\tilde a^\dag_i\tilde a_i~~
\label{bhmh}
\ee
Here $\langle i,j \rangle$ refers to all nearest neighbor pairs, $J$ is a hopping term while $U$ is an on-site repulsion proportional to ${\cal U}_0$.  The factor ${1\over N}$ is used to render ${\cal H}$ to be energy per site.

We  consider the homogeneous case and  set $\epsilon_i = 0$ at each site. This system has a superfluid-Mott insulator transition first studied by \cite{Fisher}. The insulating phase is characterized by zero compressibility and has a gap in the excitation spectrum.  The Bose-Hubbard model can describe  optical lattices and its properties, such as the excitation spectrum, had been experimentally probed \cite{Greiner, Stoferle}.

\section{Complexity of Two-Sites Bose Hubbard Model}
Hamiltonian of 2-sites Bose Hubbard Model can be written as  
\be
{\cal H}_{1,2}= {J\over 2} (\tilde a^\dag_1 \tilde a_2 +\tilde a^\dag_2 \tilde a_1) +{U\over 2}\left((\tilde a_1^\dag \tilde a_1)^2+(\tilde a_2^\dag \tilde a_2)^2-(\tilde a_1^\dag \tilde a_1)-(\tilde a_2^\dag \tilde a_2)\right)
\ee
in which we consider the periodic boundary as shown in figure 1.
\\
\\
\scalebox{0.15}{\hspace{25 cm}\includegraphics{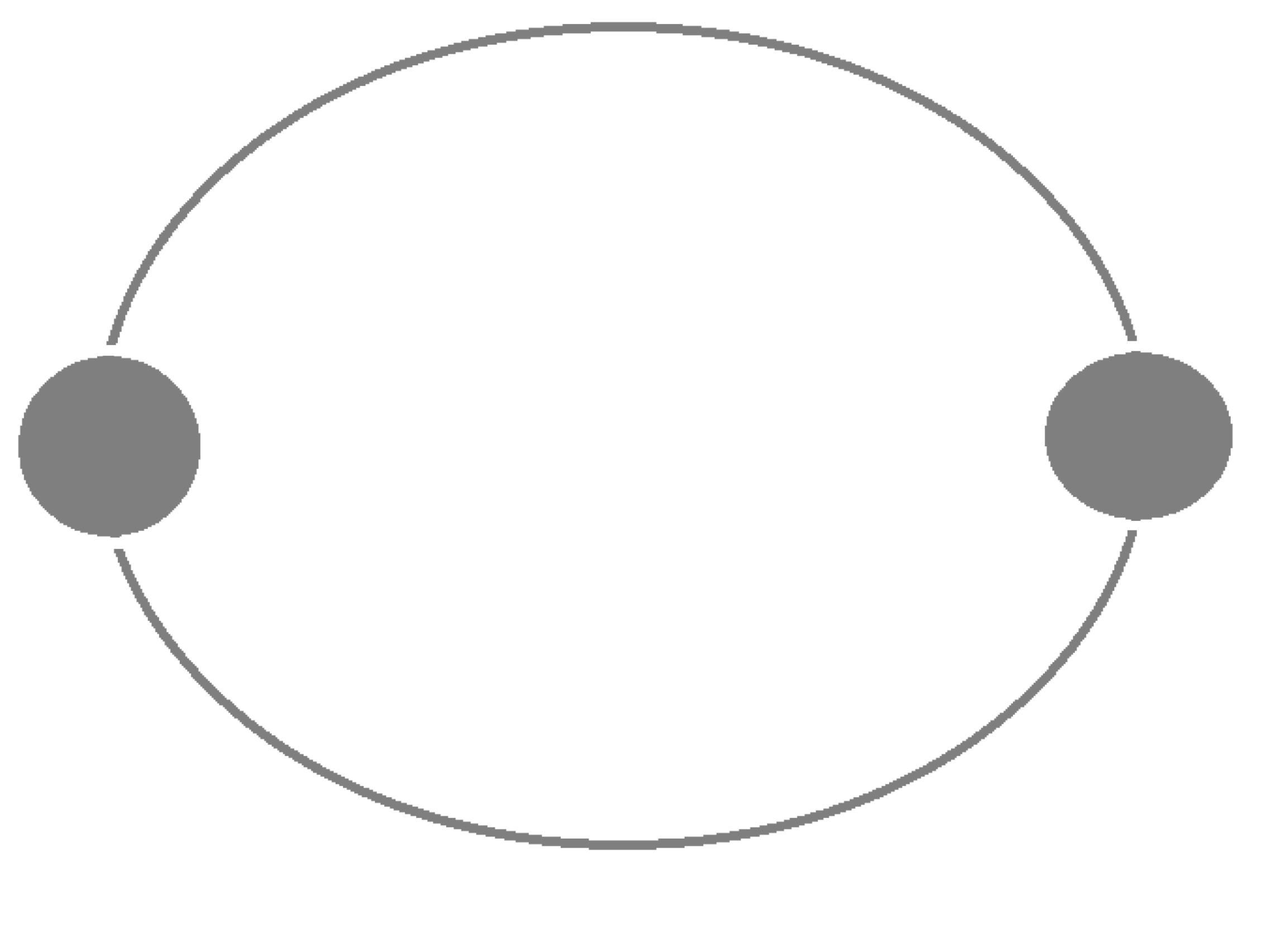}}
\\
{Figure 1:  Two sites Bose-Hubbard Model model. We choose the model with periodic boundary and there has two links. The associated complexity is plotted in figure 2.}
\\
\\
Now, we define new operators $a$ and $b$ 
\be
&&a\equiv {1\over \sqrt2}(\tilde a_1+\tilde a_2),~~b\equiv {1\over \sqrt2}(\tilde a_1-\tilde a_2)
\ee
then, by $[\tilde a_i,\tilde a_j^\dag]=\delta_{i,j}$, we have commutation relations
\be
&&[a, a^\dag]={1\over2}[\tilde a_1+\tilde a_2,\tilde a^\dag_1+\tilde a^\dag_2]={1\over2}+{1\over2}=1\\
&&[b, b^\dag]={1\over2}[\tilde a_1-\tilde a_2,\tilde a^\dag_1-\tilde a^\dag_2]={1\over2}+{1\over2}=1\\
&&[a, b^\dag]={1\over2}[\tilde a_1+\tilde a_2,\tilde a^\dag_1-\tilde a^\dag_2]={1\over2}-{1\over2}=0=[a^\dag, b]\\
&&[a, b]=[a^\dag, b^\dag]=0
\ee
Thus $\{a,a^\dag\}$ commute with $\{b,b^\dag\}$ while each set has standard  commutation relation. We can then define the number operators $n_a$ and $n_b$  
\be
 n_a&=& a^\dag a,~~~~ n_b= b^\dag b,~~~~~[n_a,n_b]=0
\ee
which define the Fock space
\be
n_a|N_a\>&=& N_a|N_a\>,~~~~~n_b|N_b\>= N_b| N_b\>
\ee
and kinematic term ${\cal H}^{(0)}$ becomes
\be
{\cal H}^{(0)}_{1,2}&=& {J\over 2} (\tilde a^\dag_1 \tilde a_2 +\tilde a^\dag_2 \tilde a_1) ={J\over 2}(a^\dag a-b^\dag b)
\ee

To proceed we begin to calculated the  interaction terms : 
\be
\tilde a_1^\dag \tilde a_2+\tilde a_2^\dag \tilde a_1&=&{1\over 2}(a^\dag+b^\dag)(a-b)+{1\over 2}(a^\dag-b^\dag)(a+b)=a^\dag a-b^\dag b=n_a-n_b\\
\tilde a_1^\dag \tilde a_1+\tilde a_2^\dag \tilde a_2&=&{1\over 2}(a^\dag+b^\dag)(a+b)+{1\over 2}(a^\dag-b^\dag)(a-b)=a^\dag a+b^\dag b=n_a+n_b
\ee
and
\be 
(\tilde a_1^\dag \tilde a_1)^2&=&\({1\over 2}(a^\dag+b^\dag)(a+b)\)^2={1\over 4}\((a^\dag a+b^\dag b) +(a^\dag b +ab^\dag)\)^2\nn\\
&=&{1\over 4}\(n_a^2+n_b^2+2n_an_b+(a^\dag b ab^\dag+ab^\dag a^\dag b)+\text{irrelevant terms}\)\nn\\
&=&{1\over 4}\(n_a^2+n_b^2+2n_an_b+(n_a(1+n_b)+(1+n_a)n_b)+\text{irrelevant terms}\)\nn\\
&=&{1\over 4}\(n_a^2+n_b^2+4n_an_b+n_a+n_b+\text{irrelevant terms}\)\\
(\tilde a_2^\dag \tilde a_2)^2&=&\({1\over 2}(a^\dag-b^\dag)(a-b)\)^2={1\over 4}\((a^\dag a+b^\dag b) -(a^\dag b +ab^\dag)\)^2\nn\\
&=&{1\over 4}\(n_a^2+n_b^2+2n_an_b-(a^\dag b ab^\dag+ab^\dag a^\dag b)+\text{irrelevant terms}\)\nn\\
&=&{1\over 4}\(n_a^2+n_b^2+2n_an_b-(n_a(1+n_b)+(1+n_a)n_b)+\text{\it irrelevant terms}\)\nn\\
&=&{1\over 4}\(n_a^2+n_b^2-n_a-n_b+\text{\it irrelevant terms}\)
\ee
in which the "{\it irrelevant terms}" are those with different  power of $a$ and $a^\dag$ and/or with different  power of  $b$ and $b^\dag$. 

In this paper we investigate the quantity $\<N_a,N_b|{\cal H}_{1.2}|N_a,N_b\>$ for the  states $|N_a,N_b\>$ with fixed $N_a$ and $N_b$, and calculate the complexity associated to the states.  In this way, only the terms that have the same power of $a$ and $a^\dag$ and that have the same power of $b$ and $b^\dag$ are {\it relevant} since that $\<N_a,N_b|{\it irrelevant\, terms}|N_a,N_b\>=0$. Therefore we neglect the {\it irrelevant terms} hereafter.

Collect all then Hamiltonian of 2-sites Bose-Hubbard model becomes
\be
{\cal H}_{1,2}&=& {J\over 2} (\tilde a^\dag_1 \tilde a_2 +\tilde a^\dag_2 \tilde a_1) +{U\over 2}\left((\tilde a_1^\dag \tilde a_1)^2-(\tilde a_1^\dag \tilde a_1)+(\tilde a_2^\dag \tilde a_2)^2-(\tilde a_2^\dag \tilde a_2)\right)\nn\\
&=&{J\over 2}(n_a-n_b)+{U\over 2}\( n_a^2+n_b^2+2 n_an_b-n_a-n_b\)+\text{\it irrelevant terms}~~~\label{H}
\ee
We will use above Hamiltonian to find the complexity of the Bose-Hubbard model. We first consider the free theory without the interaction and then the interacting theory.
\subsection{Complexity of  Free Theory}
Hamiltonian (\ref{H}) is regarded as the target system which has the kinetic term and potential term. The target kinetic term is
\be
{\cal K}^{\rm (tar)}_{a,b}= {J\over 2} (\tilde a^\dag_1\tilde  a_2 +\tilde a^\dag_2 \tilde a_1)={J_a\over 2}\,a^\dag a-{J_b\over 2}\,b^\dag b
\ee
in which, for later convenience, we define
\be
J_a=J_b=J
\ee
To calculate the complexity we choose the following kinetic term  as  that of the reference state
\be
K^{\rm (ref)}_{a^{\rm (ref)},\,b^{\rm (ref)}}&=&{J_f\over 2}(a^{\rm (ref)})^\dag \,a^{\rm (ref)}  -{J_f\over 2}\, (b^{\rm (ref)})^\dag\, b^{\rm (ref)}
\ee
where ${J_f}$  is the coupling strength of the reference state.

Now we see that with the replacement 
\be
a^{\rm (ref)}\rightarrow \sqrt {J_a\over J_f}\,a,~~b^{\rm (ref)}\rightarrow \sqrt {J_b\over J_f}\,b~~~\label{rep}
\ee
one can obtain $K^{\rm (tar)}$ from  $K^{\rm (ref)}$, i.e.
\be
 K^{\rm (\rm ref)}\rightarrow  K^{\rm (tar)} 
\ee
In the operator approach the gate matrix defined in (\ref{gate}) is constructed by the transformation from target operator to reference operator in above relation. 
\subsubsection{Ground State}
Consider first the ground state.  Target ground state is annihilated by  $a,b$, i.e. $a|0,0\>=b|0,0\>=0$, and   reference ground state is annihilated by  $a^{\rm (ref)},b^{\rm (ref)}$, i.e.  $a^{\rm (ref)}|0,0\>_{\rm ref}=b^{\rm (ref)}|0,0\>_{\rm ref}=0$. In the operator approach to complexity the gate matrix $U(s)$, defined in   (\ref{gate}), connecting the target operator to  reference operator can be read from  (\ref{rep}) 
\be
\left(
\ba{cc}
a&b\\
\ea\right)
=U(1)
\left(
\ba{c}
a^{\rm (ref)}\\
b^{\rm (ref)}\\
\ea\right)~~~\label{U},~~~~U(1)=\left(
\ba{cc}
\sqrt {J_a\over J_f}&0\\
0&\sqrt {J_a\over J_f}\\
\ea\right)
\ee
in which the initial condition is $U(0)=diag(1,1)$.  Since the  transformation matrix $U(1)$ is diagonal we can choose  ${\cal O_I}=1$ in (\ref{gate}) and  have a simple relation
\be
U(1)=\left(
\ba{cc}
e^{\int_0^1\,ds \,Y_a(s)}&0\\
0&e^{\int_0^1\,ds \,Y_b(s)}\\
\ea\right)
=\left(
\ba{cc}
\sqrt {J_a\over J_f}&0\\
0&\sqrt {J_a\over J_f}\\
\ea\right)
\ee
The associated solutions of $Y_{(1,2)}(s)$  are
\be
Y_a(s)&=&\ln \left(\sqrt {J_a\over J_f}\right),~~~Y_b(s)=\ln \left(\sqrt {J_b\over J_f}\right)~~~~\label{Y}
\ee
which satisfied the initial condition.  For the convention used  in later we define new functions 
\be
R^{(0)}_a={J_a\over J_f},~~~R^{(0)}_b={J_b\over J_f},~~~~R^{(0)}={J\over J_f}~~~~~\label{R0}
\ee
in which $(0)$ means a quantity of the zero order of interaction.  The squared distance for ground state, denoted as $(D^{(0)}_{(0,0)})^2$, between target and reference state can be calculated by formula  (\ref{distance})
\be
(D^{(0)}_{(0,0)})^2=Y_a(1)^2+Y_b(1)^2=\left(\ln \left(R^{(0)}_a\right)\right)^2+\left(\ln \left(R^{(0)}_b\right)\right)^2~~~~\label{d0}
\ee
in which $(0,0)$ means a quantity of the ground state of two sites system.  This formula has a similar form as that in scalar field theory \cite{Jefferson1707, Huang1905}.
\subsubsection{Excited States}
Consider next the $\{N_a^{th},\,N_b^{th}\}$ excited state which is defined by 
\be
|N_a,N_b\>={(a^\dag)^{N_a+1}(b^\dag)^{N_b+1}\over \sqrt{(N_a+1)!\,(N_b+1)!}}|0,0\>,~~~a^{N_a+1}\,b^{N_b+1}|N_a,N_b\>=0
\ee
 In the operator approach to complexity the gate matrices can be  read from the transformations 
\be
\overbrace{a^{\rm (ref)}\,....a^{\rm (ref)}}^{N_q+1}\,\overbrace{b^{\rm (ref)}\,....b^{\rm (ref)}}^{N_b+1}\,\,\Rightarrow\,\,\overbrace{\sqrt {J_a\over J_f}\,a\,...\sqrt {J_a\over J_f}\,a}^{N_a+1}\,\overbrace{\sqrt {J_b\over J_f}\,b\,....\sqrt {J_b\over J_f}\,b}^{N_b+1}\,
\ee
Then, the gate matrix connecting the target operator with  referenct operator  in  (\ref{gate}) becomes a $(N_a+1)\times (N_b+1)$ diagonal matrix $U(s)$ 
\be
U(1)=\text{Diag}\left(\overbrace{\sqrt {J_a\over J_f},...,\sqrt {J_a\over J_f}}^{N_a+1}\,,\overbrace{\sqrt {J_b\over J_f},...,\sqrt {J_b\over J_f}}^{N_b+1}\right)
\ee
which redues to  (\ref{U}) in the case of ground state $N_a=N_b=0$. 

Follow the discussions in before the gate matrix $U_i$ defined in  (\ref{gate}) now becomes
\be
U_i(1)&=&\sqrt {J\over J_f}\,,~~{\rm with}~~U_{i}(0)=1,~~1\le i\le N_2+N_1+2
\ee
 after use the relation $J_a=J_b=J$.  The associated functions of $Y_i(s)$ in   (\ref{gate}) are
\be
Y_i(s)&=&\ln\(\sqrt {J\over J_f}\),~~1\le i\le N_a+N_b+2
\ee
The squared distance for excited  state, denoted as $D^2_{(N_a,N_b)}$, between target and reference state calculated from  (\ref{distance}) is
\be
(D^{(0)}_{(N_a,N_b)})^2&=&\sum_{i=1}^{N_a+N_b+2}Y_i(1)^2=(N_a+N_b+2)\left(\ln \(\sqrt{R^{(0)}}\)\right)^2,~~~R^{(0)}={J\over J_f}~~\label{2D}
\ee
This has a similar form as the result of scalar theory obtained earlier in \cite{Huang1905}. 

 Since that  state wavefunction is described by $\Psi_n(x)={1\over \sqrt{n!}}\<x|(a^\dag)^n|0\>$  the gate matrix of wavefunction, $\Psi_n(x)$, is thus related to the gate matrix of  field operators, $(a^\dag)^n$.
\subsection{Complexity of Interacting Theory}
We next study the correction to the complexity due to the interaction term.  We write the Hamiltonian as the summation of kinetic (free) term and potential (interaction) term 
\be
{\cal H}_{1,2}&=& {J\over2} (\tilde a^\dag_1 \tilde a_2 +\tilde a^\dag_2 \tilde a_1) +{U\over 2}\left((\tilde a_1^\dag \tilde a_1)^2-(\tilde a_1^\dag \tilde a_1)+(\tilde a_2^\dag \tilde a_2)^2-(\tilde a_2^\dag \tilde a_2)\right)\nn\\
&=&{J\over2}(n_a-n_b)+{U\over 2}\( n_a^2+n_b^2+2 n_an_b-n_a-n_b+\text{irrelevant terms} \)\nn\\
&=&{\cal K}_{a,b}^{\rm (tar)}+{\cal V}_{a,b}
\ee
 The interacting  term is
\be
{\cal V}_{a,b}^{\rm (tar)}&=&{U\over 2}\( n_a^2+n_b^2+2 n_an_b-n_a-n_b \)~~~\label{2V}
\ee
To  consider $\<N_a,N_b|{\cal V}_{a,b}|N_a,N_b\>$ for the excited state $|N_a,N_b\>$  we write above equation as
\be
{\cal V}_{a,b}^{\rm (tar)}&=&{U\over 2}\(N_a\,a^\dag a+N_b\,b^\dag b+N_a\,b^\dag b+N_b\,a^\dag a-a^\dag a-b^\dag b \)
\ee
 Therefore
\be
{\cal H}_{a,b}^{\rm (tar)}
={J\over2}\left(1-{U\over J}\,(N_a+N_b-1\)\right) a^\dag  a+{J\over2}\left(-1-{U\over  J}\,(N_a+N_b-1\)\right) b^\dag  b~~\label{VN=2a}
\ee
The associated  Hamiltonian of the reference state  is   
\be 
H^{\rm (ref)}_{a^{\rm (ref)},\,b^{\rm (ref)}}&=&{J_f\over2}\left(1-{U\over J_f}\,(N_a-1\)\right) (a^{\rm (ref)})^\dag  a^{\rm (ref)}+{J_f\over2}\left(-1-{U\over J_f}\,(N_b-1\)\right) (b^{\rm (ref)})^\dag  b^{\rm (ref)}~~\label{VN=2b}\nn\\
\ee
Above choice  satisfies a desirable property of the reference state that it does not contain any entanglement between operators $\{N_a,\, a^{\rm (ref)},\,  (a^{\rm (ref)})^\dag\} $ and $\{N_b,\, b^{\rm (ref)},\,  (b^{\rm (ref)})^\dag\} $. The property  in operator  approach is like as that in coordinate approach in which a desirable property of the reference state is that it should not contain any entanglement between the original coordinates $x_1$ and $x_2$,  as discussed  in  \cite{Jefferson1707, Huang2008}. 
\subsubsection{First Order}
In the case of zero-order of $U$ the Hamiltonian  remain only the kinetic term.  This is the free case discussed in previous section. Now consider the perturbation to the complexity for the two-sites Bose-Hubbard Model.   At the first order of $U$ we can use (\ref{VN=2a}) and  (\ref{VN=2b}) to find transformations
\be
\left\{\ba{ccc}
J\left(1-{U\over J}\,(N_a+N_b-1)\right) a^\dag  a&\rightarrow&J_f\left(1-{U\over J_f}\,(N_a-1)\right) (a^{\rm (ref)})^\dag a^{\rm (ref)}\\
\\
J\left(1+{U\over J}\,(N_a+N_b-1)\right) b^\dag  b&\rightarrow&J_f\left(1+{U\over  J_f}\,(N_b-1)\right) (b^{\rm (ref)})^\dag  b^{\rm (ref)} 
\ea  \right.    ~~~\label{I0}
\ee
To proceed we notice that the factors $N_{(a,b)}$ are within the coupling term, i.e. ${U\over 2J}$ and we only need to consider their zero-order transform. Thus, by (\ref{R0}), we  can put following replacement  
\be
N_a\rightarrow R^{(0)}_a\,N_a\\
N_b\rightarrow R^{(0)}_b\,N_b
\ee
into (\ref{I0}) and final formulas in first-order transformations are
\be
\left\{\ba{ccc}
R^{(1)}_a{(N_a,\,N_b)}&=&{J\left(1-{U\over J}\,\left(R^{(0)}_aN_a+R^{(0)}_bN_b-1\right)\right)\over J_f\left(1-{U\over J_f}\,(N_a-1)\right)}\\
R^{(1)}_b{(N_a,\,N_b)}&=&{J\left(1+{U\over J}\,\left(R^{(0)}_aN_a+R^{(0)}_bN_b-1\right)\right)\over J_f\left(1+{U\over J_f}\,(N_b-1)\right) }~~~\label{R21}\ea  \right.   ~~~\label{R2}
\ee
The  first-order square distance is
\be
\(D^{(1)}_{(N_a,N_b)}\)^2=(N_a+1)\left(\ln \(\sqrt {R^{(1)}_a(N_a,N_b)}\)\right)^2+(N_b+1)\left(\ln \(\sqrt {R^{(1)}_b(N_a,N_b)}\)\right)^2 ~~\label{2siteR}
\ee
which reduce to the square distance formula of free theory in (\ref{d0}) for ground (i.e.$N_a=N_b=0$)
\subsubsection{n'th-Order}
Extending to higher-order interactions is straightforward.  In terms of new variables
\be
g={U\over J},~~~~\gamma={J\over J_f}~~~\label{UJ}
\ee
the recursion relations are 
\be
\left\{\ba{ccc}
R^{(n)}_a{(N_a,\,N_b)}&=&{\gamma\left(1-{g}\,\left(R^{(n-1)}_aN_a+R^{(n-1)}_bN_b\,-1\right)\right)\over \left(1-{g\,\gamma}\,(N_a-1)\right)},~~~~~~~~~~R^{(0)}_a={\gamma}\\
R^{(n)}_b{(N_a,\,N_b)}&=&{\gamma\left(1+{g}\,\left(R^{(n-1)}_aN_a+R^{(n-1)}_bN_b\,-1\right)\right) \over J_f\left(1+{g\,\gamma}\,(N_b-1)\right) },~~~~~~~~~~R^{(0)}_b={\gamma}\ea  \right.  ~~\label{R2n}  
\ee
with initial values  $R^{(0)}_{(a,b)}$ defined in (\ref{R0}) with  $J_a=J_b=J$.  
For excited states, the $n$-order square distance is
\be
\(D^{(n)}_{(N_a,N_b)}\)^2=(N_a+1)\left(\ln \(\sqrt {R^{(n)}_a(N_a,N_b)}\)\right)^2+(N_b+1)\left(\ln \(\sqrt {R^{(n)}_b(N_a,N_b)}\)\right)^2~~\label{D2n}  
\ee 
which is the  general formula of $n'th$-order complexity of two-sites Bose-Hubbard model.  
\subsection{Complexity and Quantum Phase Transition}
We now use above formula to investigate the quantum phase transition in two-sites Bose-Hubbard model.  Consider first the ground in which $N_a=N_b=0$. Then (\ref{R2n})  become 
\be
R^{(n)}_a{(0,\,0)}&=&{\gamma \left(1+g\right)\over 1+g\,\gamma},~~~~~~~R^{(n)}_b{(0,\,0)}={\gamma \left(1-g\right)\over1-g\,\gamma}~~~~\label{ER0}
\ee 
in which $g={U\over J},\,\gamma={J\over J_f}$, defined in (\ref{UJ}), describes the ratio of one-site repulsion and hopping constant. We expect that complexity will be divergent at critical value $g_c$, which indicates the quantum phase transition occurs at this point.
\\

To proceed we make following comments about above relation :

1.  $R^{(n)}_{a,b}{(0,\,0)}$ are the qualities at n'th order of interaction strength  g. Thus  (\ref{ER0}) gives
\be
R^{(0)}_{a,b}{(0,\,0)}={J\over J_f}=\gamma 
\ee
 which are those in (\ref{R0}).  To first order (\ref{ER0}) gives
\be
R^{(1)}_a{(0,\,0)}&=&\gamma +\gamma (1-\gamma) g,~~~~~~~R^{(1)}_b{(0,\,0)}=\gamma -\gamma (1-\gamma) g
\ee
and using (\ref{ER0}) we can find  any order of $R^{(n)}_{a,b}{(0,\,0)}$ by series expansion of  $g$. 

2. In fact the formulas (\ref{ER0}) are the exact values and we can let $n\to\infty$. Note that being able to obtain the exact formula for ground state complexity is a special property in our method.  Now we see that  $R^{(n\to\infty)}_b{(0,\,0)}$  in (\ref{ER0}) becomes zero at critical value $g=g_c=1$.  This implies that the associated complexity, $(D^{(n\to\infty)}_{(0,0)})^2$, is divergent and thus it is the quantum phase transition point. 

3. We plot the figure 2 to explicitly see this property.
\\
\\
\scalebox{.5}{\hspace{7cm}\includegraphics{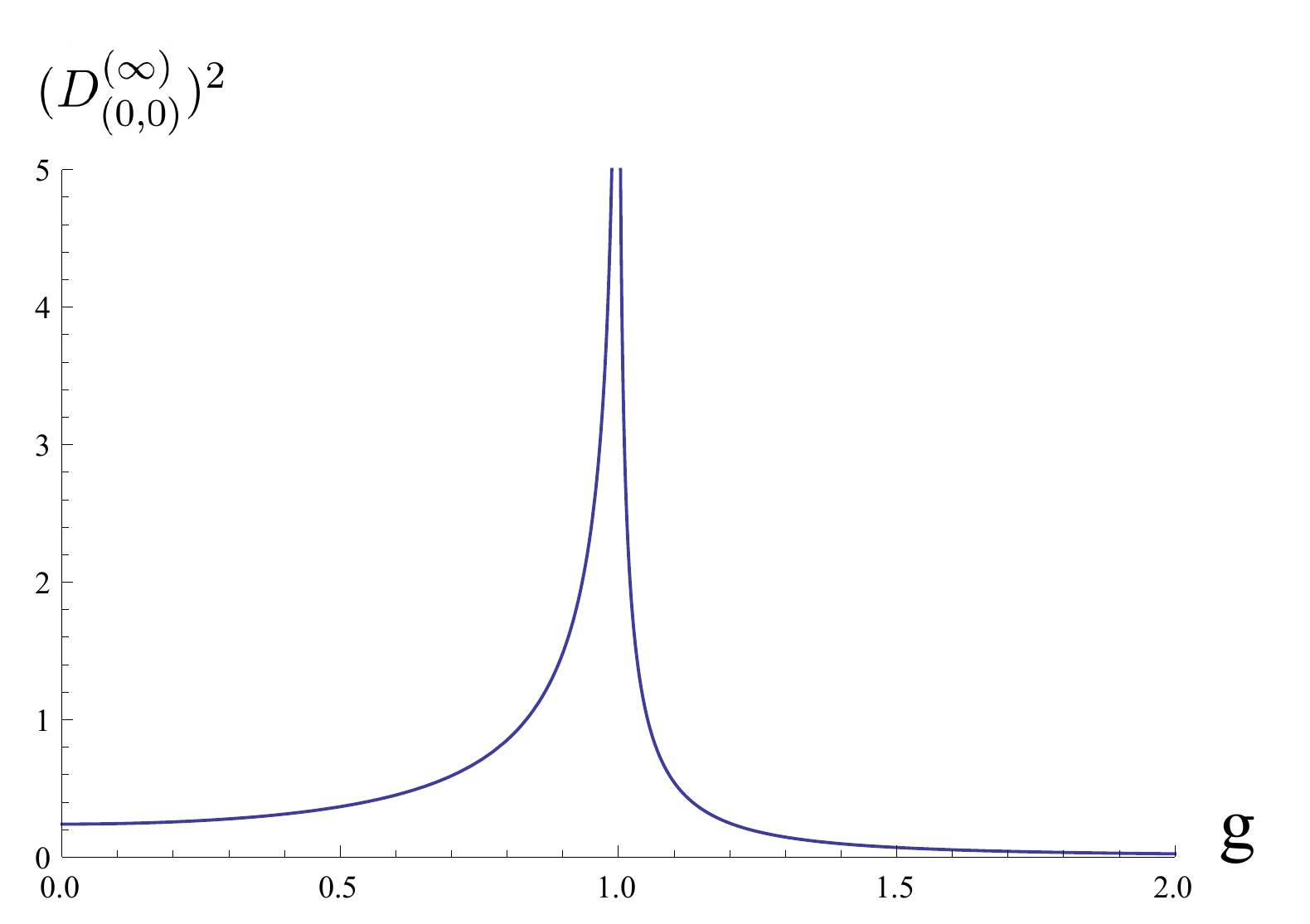}}
\\
{Figure 2:  Dependence of complexity $(D^{(n\to\infty)}_{(0,0)})^2$ on  $``g"$ in 2-sites Bose-Hubbard model. The divergence at $g=g_c=1$ indicates the quantum phase transition at this point. The figure is plotted with $\gamma_L=1/2$ while $\gamma_R=3/2$.}
\\

4.  The value of $g$ we considered  is positive. Thus  the factor $(1-g\,\gamma)$ in formulas (\ref{ER0}) could become zero at $g_0= \gamma^{-1}$ which depends on the reference state parameter $\gamma$.   To avoid  $"g_0"$ become a critical point the  following  scheme is proposed.  For range of $g>g_c$ we choose $\gamma=\gamma_R>1$ while for range of $g<g_c$ we choose $\gamma=\gamma_L<1$.  Under this prescription the value of $(1-g\,\gamma)$ is always positive and it remains only one  critical point at $g=g_c$.

5. Notice that, in this way, the complexity calculated  for  $g>g_c$ is that with $\gamma=\gamma_R>1$ while the complexity calculated with $g<g_c$ is that with $\gamma=\gamma_L<1$.  Therefore, one could compare the magnitude of complexity between  the systems of   $g>g_c$.  One could also compare the magnitude of complexity between  the systems of   $g<g_c$.  However, it makes no sense to compare the magnitude of complexity between  the systems of   $g<g_c$  and  of systems  $g>g_c$ since  the reference state parameter $\gamma$ is different between them.   The figure 2 is plotted with $\gamma_L=1/2$ while $\gamma_R=3/2$.

6. Consider next  the excited ground in which $N_a=N_b=1$. For $n\ge1$ (\ref{R2n})  becomes 
\be
R^{(n)}_a{(1,\,1)}&=&\gamma \left(1+g(1-2\gamma)\right),~~~~~~~R^{(n)}_b{(1,\,1)}=\gamma \left(1-g(1-2\gamma)\right),~~~n\ge1~~~~\label{ER1}
\ee 
As that in ground case,  the formulas (\ref{ER1}) are the exact values and we can let $n\to\infty$. We also see that either $R^{(n\to\infty)}_a{(1,\,1)}$ or $R^{(n\to\infty)}_a{(1,\,1)}$ becomes zero at critical value $g=g^{(N=1)}_c= {\pm1\over 1-2\gamma}$.  This implies that the associated complexity, $(D^{(n\to\infty)}_{(1,1)})^2$, is divergent at the critical value and thus it is the quantum phase transition point. Note that the critical point is a physical property and it shall not depend on the reference state parameter $\gamma$. However,  the calculated value $g^{(N=1)}_c$  depends  on $\gamma$. This feels short of author expectation.  The reason may be that the model described in (\ref{bhmh}) is that keeps only the lowest vibrational state and could only describe the ground state property.  The analysis in above, however, have shown a possible way to calculate excited-state complexity in other many-body models.
\section{Complexity of N-Sites Bose-Hubbard Model}
Hamiltonian of N-sites closed chain of  Bose-Hubbard model defined in (\ref{bhmh}) can be written as  
\be
{\cal H}_N = {J\over 2N} \sum_{j=0}^N (\tilde a^\dag_{j+1}\tilde a_j+\tilde a^\dag_{j-1}\tilde a_j)+ \frac{U}{N}\sum_{j=0}^N \tilde n_j (\tilde n_j - 1),~~~~~~~\tilde n_j=\tilde a^\dag_j\tilde a_j~~ \label{bhmhN} 
\ee
in which we consider the periodic boundary as shown in figure 3.
\\
\\
\scalebox{0.15}{\hspace{25 cm}\includegraphics{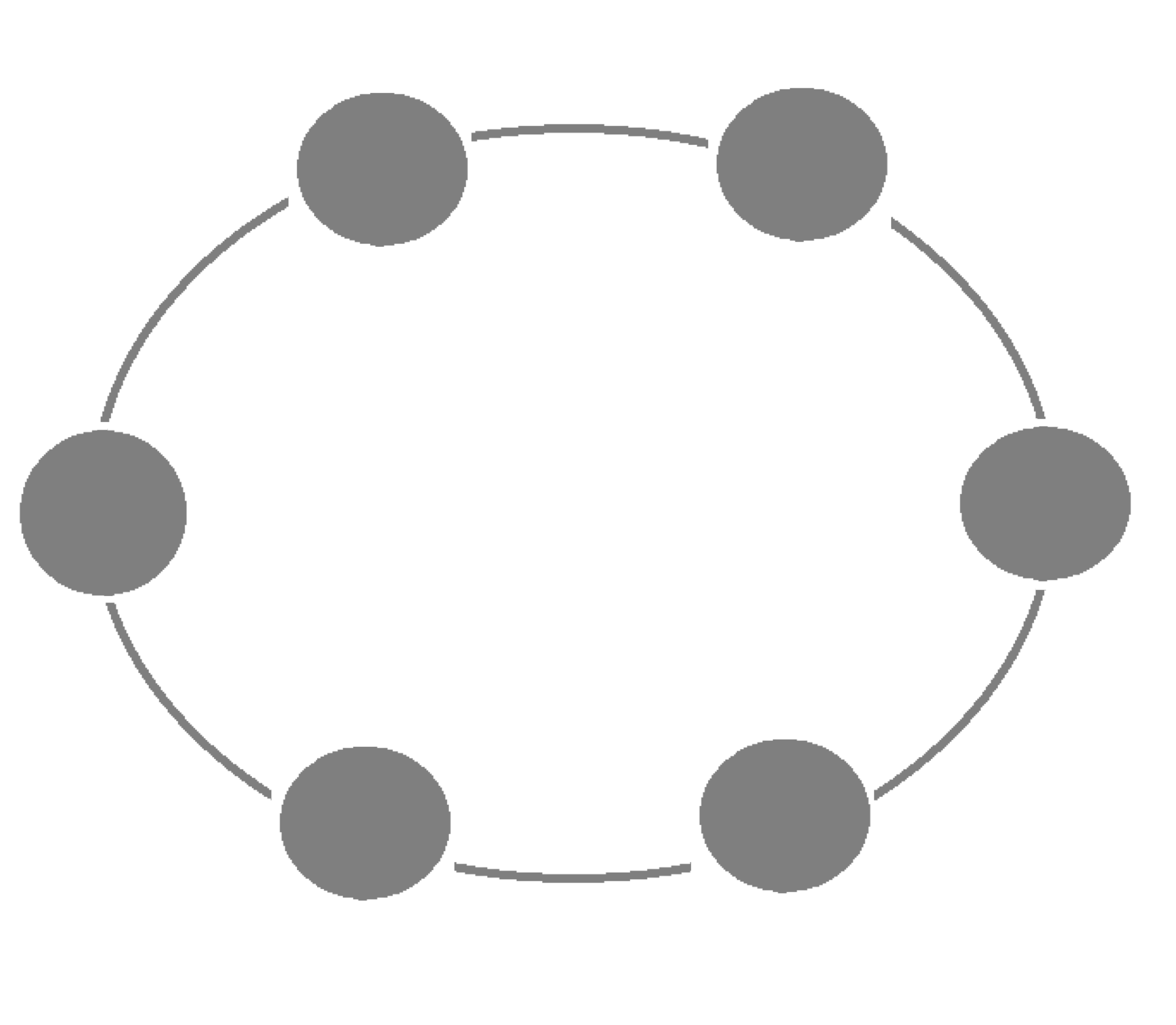}}
\\
{Figure 3:  Six-sites Bose-Hubbard Model model. We choose the model with periodic boundary and there has six links. The associated complexity is plotted in figure 4.}
\\
\\
We can generalize method in previous section to define the new operators
\be
a_k&=& {1\over \sqrt N}\sum_{j=0}^{N-1}\,exp\Big({2\pi i k\over N} \,j\Big)\,\tilde a_j,~~ a^\dag_k= {1\over \sqrt N}\sum_{j=0}^{N-1}\,exp\Big({-2\pi i k\over N} \,j\Big)\,\tilde a^\dag_j\\
\tilde a_k&=& {1\over \sqrt N}\sum_{j=0}^{N-1}\,exp\Big({-2\pi i k\over N} \,j\Big)\, a_j,~~
\tilde a^\dag_k= {1\over \sqrt N}\sum_{j=0}^{N-1}\,exp\Big({2\pi i k\over N} \,j\Big)\, a^\dag_j
\ee
in which we impose a periodic boundary condition $\tilde a_{\rm k+N+1}=\tilde a_k$ and  $\tilde a^\dag_{\rm k+N+1}=\tilde a^\dag_k$.

 Note that the  relative sign between the Fourier series of $a_k$ and $a^\dag_k$ is important to have standard commuation relation \cite{Jefferson1707}. Then, above  definitions leads to  simple relations  
\be
~[\tilde a_i,\tilde a_j]=[\tilde a^\dag_i,\tilde a^\dag_j]=0,~~[\tilde a_i,\tilde  a^\dag_j]=\delta_{i,j}~~~\Rightarrow~~~[a_i,a_j]=[a^\dag_i,a^\dag_j]=0,~~[a_i,a^\dag_j]=\delta_{i,j}
\ee
and
\be
[a_j,a^\dag_k]&=&{1\over N}\sum_{m,\ell}\,e^{{2\pi\,i (m\,j-k\,\ell)\over N}}\,[a_m,a^\dag_\ell]={1\over N}\sum_{m,\ell}\,e^{{2\pi\,i (m\,j-k\,\ell)\over N}}\,\delta_{m,\ell}={1\over N}\sum_{\ell}\,e^{{2\pi\,i \ell(m-k)\over N}}=\delta_{j,k}~~~~~
\ee
The general states can then be constructed by creation operators $a_k^\dag$ 
\be
|{ N}_0,\,{ N}_1\cdot\,\cdot\,\cdot\,{ N}_{N-1}\>=a_0^{{N}_0}\,a_1^{{ N}_1}\cdot\,\cdot\,\cdot\,a_{N-1}^{{ N}_{N-1}}\,|0\>
\ee
due to the commutation relation between the new operators.
\\

We now express N-sites Hamiltonian ${\cal H}_N$ in terms of new operators $a_j$ and $ a^\dag_j$.  Kinematic term is
\be
\sum_j\tilde a^\dag_{j+1}\tilde a_j+\sum_j\tilde a^\dag_{j-1}\tilde a_j&=&{1\over N}\sum_j\sum_{m,k}e^{{2\pi\,i (-m(j+1)+kj)\over N}}\,a^\dag_m  a_k+{1\over N}\sum_j\sum_{m,k}e^{{2\pi\,i (-m(j-1)+kj))\over N}}\,a^\dag_m  a_k\nn\\
&=&\sum_ke^{{-2\pi\,i k\over N}}\,a^\dag_k  a_k+\sum_ke^{{2\pi\,i m\over N}}\,a^\dag_k  a_k\\
&=&2\sum_{k=0}^{N-1}\,\cos\left({2\pi k\over N}\right)\,a^\dag_k  a_k
\ee
The kinematic term in   ${\cal H}_N$ can thus be expressed as
\be
{\cal K}_N={J\over2N}\left(\sum_j\tilde a^\dag_{j+1}\tilde a_j+\sum_j\tilde a^\dag_{j-1}\tilde a_j\right)&=&{J\over N}\sum_{k=0}^{N-1}\cos\left({2\pi k\over N}\right)\,a^\dag_k  a_k
\ee
which reduces to the formula of  N=2 case in previous section. The interaction terms has two parts. One is
\be
\sum_j \tilde n_j=\sum_j\tilde a^\dag_j\tilde a_j={1\over N}\sum_j\sum_{m,k}e^{{2\pi\,i (-m+k)j\over N}}\,a^\dag_m a_k=\sum_{m,k}\delta_{m,k}\,a^\dag_m  a_k=\sum_k n_k
\ee
Another one is
\be
\sum_j \tilde n_j^2&=&\sum_j \tilde a^\dag _j\tilde a _j\tilde a^\dag _j\tilde a _j={1\over N^2}\sum_j\sum_{\alpha,\beta,\gamma,\delta}e^{2\pi j(-\alpha+\beta-\gamma+\eta)\over N} a^\dag _\alpha a _\beta a^\dag _\gamma a_\eta\\
&=&{1\over N}\sum_{\alpha,\beta,\gamma,\eta} a^\dag_\alpha a_\beta a^\dag _\gamma a_\eta\,\delta_{\alpha-\beta+\gamma-\eta,0}
\ee
The delta function in here means that once $\alpha=\beta$ then $\gamma=\eta$ and, if $\alpha\ne\beta$ then $\gamma\ne\eta$. In the first case we have term
\be
{1\over N}\sum_{\alpha,\gamma} a^\dag_\alpha a _\alpha a^\dag _\gamma a_\gamma={1\over N}\sum_{\alpha,\gamma} n _\alpha \,a^\dag _\gamma a_\gamma
\ee
In the second  case, following the discuss in previous section we have to choose the {\it relevant term} which is
\be
{1\over N}\sum_{\alpha\ne\gamma} a^\dag_\alpha a _\gamma a^\dag _\gamma a_\alpha&=&{1\over N}\sum_{\alpha\ne\gamma} n _\alpha \,a_\gamma a^\dag _\gamma ={1\over N}\sum_{\alpha\ne\gamma} n _\alpha \,(1+a^\dag _\gamma a_\gamma)\nn\\
&=&\left({N-1\over N}\sum_\alpha  n _\alpha\right)+{1\over N}\sum_{\alpha\ne\gamma} n _\alpha a^\dag _\gamma a_\gamma
\ee
Collect all and the interaction  term in   ${\cal H}_N$ becomes
\be
{\cal V}_N&=&{U\over N}\(\sum_{\alpha,\gamma} n _\alpha a^\dag _\gamma a_\gamma+\sum_{\alpha\ne\gamma} n _\alpha a^\dag _\gamma a_\gamma-\sum_\alpha  n _\alpha\)+\text{\it irrelevant terms}\nn\\
&=&{U\over N}\(\sum_{k=0}^{N-1} n _k a^\dag _k a_k+2\sum_{(k,j)=0}^{N-1} n _j a^\dag _k a_k-\sum_{k=0}^{N-1}  n _k\)+\text{\it irrelevant terms}
\ee
where $(k,j)$ denotes any pair at site $k$ and site $j$.  When N=2 above formula  reduces to that of 2-sites formula  in (\ref{2V}).
\subsection{Complexity of  Free Theory}
Let us first discuss the case of free theory. Following the scheme detailed in previous section the  target system kinematic term and  reference  system kinematic term are
\be
{\cal K}^{\rm (tar)}_N&=&{J\over N}\sum_{k=0}^{N-1}\cos\left({2\pi k\over N}\right)\,a^\dag_k \, a_k~~\label{Jk}\\
{\cal K}^{\rm (ref)}_N&=&{J_f\over N}\sum_{k=0}^{N-1}\cos\left({2\pi k\over N}\right) a_k^{\rm (ref)\dag}\, a_k^{\rm (ref)}~~\label{Jf}
\ee
where $J_f$  is the coupling strength of the reference state.

Now we see that with the replacement 
\be
a^{\rm (ref)}_k\rightarrow \sqrt {J\over J_f}\,a_k
\ee
one can obtain $K^{\rm (tar)}$ from  $K^{\rm (ref)}$, i.e.
\be
 K^{\rm (\rm ref)}\rightarrow  K^{\rm (tar)} 
\ee
As mentioned before, in the operator approach the gate matrix defined in (\ref{gate}) is constructed by the transformation from target operator to reference operator in above relation.  Thus the complexity, 
square distance,  for the  general excited state $\{{ N}_0,{ N}_1\cdot\cdot\cdot,{ N}_{N-1}\}$ is easy to be found by the method discussed in the 2-site case. The result is 
\be
D^2_{\{{N}_0,{ N}_1\cdot\cdot\cdot,{ N}_{N-1}\}}&=&\sum_{k=0}^{N-1}\,({N}_k+1)\left(\ln \(\sqrt{ R^{(0)}_k}\)\right)^2,~~~~~~R^{(0)}_k={J\over J_f}~~~\label{NR0}
\ee
which is a simple extension of 2-sites formula in  (\ref{2D}).
\subsection{Complexity of Interacting Theory}
To include the interaction we first  collect the total Hamiltonian 
\be
{\cal H}_N&=&{1\over N}\sum_{k=0}^{N-1}J\cos\left({2\pi k\over N}\right)\,a^\dag_k  a_k+{U\over N}\(\sum_{k=0}^{N-1} n _k a^\dag _k a_k+2\sum_{(k,j)=0}^{N-1} n _j a^\dag _k a_k-\sum_{k=0}^{N-1}  a^\dag _k a_k\)\\
&=&{1\over N}\sum_{k=0}^{N-1}\left(J\cos\left({2\pi k\over N}\right)\,+{U}\(-1+ n _k+\sum_{j\ne k}^{N-1} n _j\) \right)a^\dag _k a_k
\ee
where $ n _k=a^\dag _k a_k$ is the number operator. Then, following the scheme described in previous section we furthermore write the total Hamiltonian of target state and reference state as
\be
{\cal H}^{\rm (tar)}_N&=&{1\over N}\sum_{k=0}^{N-1}\left(J \cos\left({2\pi k\over N}\right)\,+{U}\(-1+ { N} _k+\sum_{j\ne k}^{N-1} {N} _j\)\right) a^\dag _k a_k\\
{\cal H}^{\rm (ref)}_N&=&{1\over N}\sum_{k=0}^{N-1}\left(J_f \cos\left({2\pi k\over N}\right)\,+{U}\left(-1+ { N} _k\right)\right)a_k^{\rm (ref)\dag}\, a_k^{\rm (ref)}
\ee
where ${N} _k$ is an integral number defined by  $ n _k|0\>=|{ N} _k\>$. The reference operator in above satisfies a desirable property of the reference state that it does not contain any entanglement between operators $a_k^{\rm (ref)} $ and $a_j^{\rm (ref)}$ if $k\ne j$,  as discussed  in  \cite{Jefferson1707, Huang2008}.

\subsubsection{First Order}
In the case of zero-order of $U$ the Hamiltonian  remain only the kinetic term.  This is the free case. Now consider the perturbation to the complexity for the N sites Bose Hubbard Model.   At the first order of $U$ the  transformation  from target operator to reference operator is chosen as following
\be
\left(J \cos\left({2\pi k\over N}\right)\,+{U}\(-1+ { N} _k+\sum_{j\ne k}^{N-1} {N} _j\)\right) a^\dag _k a_k&\rightarrow& 
\left(J_{f} \cos\left({2\pi k\over N}\right)\,+{U}\left(-1+ { N} _k\right)\right)a_k^{\rm (ref)\dag}\, a_k^{\rm (ref)}~~\label{V1}\nn\\
\ee
Now, we notice that the factors $N_k,\,N_j$ are within the coupling term, i.e. ${U}$, and we only need to consider their zero-order transform. Thus, by (\ref{NR0}), we  shall put following replacement  
\be
N_k\rightarrow R^{(0)}_k\,N_k
\ee
into (\ref{V1}) and final formulas in first-order transformation is
\be
R^{(1)}_k({\{N_i\}})&=&{J \cos\left({2\pi k\over N}\right)\,+{U}\(-1+ R^{(0)}_k\,{ N} _k+\sum_{j\ne k}^{N-1} {N} _j\)\over J_{f} \cos\left({2\pi k\over N}\right)\,+{U}\left(-1+ { N} _k\right)}    
\ee
Therefore the  first-order complexity or square distance is
\be
\(D^{(1)}_{\{N_i\}}\)^2=\sum_{k=0}^{N_1}(N_k+1)\,\left(\ln \(\sqrt {R^{(1)}_k({\{N_i\}})}\)\right)^2
\ee
The case of ${\{N_i\}}=\{N_a,N_b\}$ above formula reduce to the square distance formula of 2-site theory in (\ref{2siteR}).
\subsubsection{n'th-Order}
Extending to higher-order interactions is straightforward. The recursion relation  is 
\be
R^{(n)}_k({\{N_i\}})&=&{J \cos\left({2\pi k\over N}\right)\,+{U}\(-1+ R^{(n-1)}_k\,{ N} _k+\sum_{j\ne k}^{N-1} {N} _j\)\over J_{f} \cos\left({2\pi k\over N}\right)\,+{U}\left(-1+ { N} _k\right)}\\
&=&{\gamma \left(\cos\left({2\pi k\over N}\right)\,+{g}\left(-1+ R^{(n-1)}_k\,{ N} _k+\sum_{j\ne k}^{N-1} {N} _j\right)\right)\over  \cos\left({2\pi k\over N}\right)\,+{\gamma\,g}\left(-1+ { N} _k\right)}~~\label{RNn}    
\ee
with initial values  $R^{(0)}_{(a,b)}$ defined in (\ref{NR0}) and  the $n$-order square distance is
\be
\(D^{(n)}_{\{N_i\}}\)^2=\sum_{k=0}^{N_1}(N_k+1)\,\left(\ln \(\sqrt {R^{(n)}_k({\{N_i\}})}\)\right)^2~~\label{FF}
\ee
These complete our derivations.
\subsection{Complexity and Quantum Phase Transition}
Consider first the ground in which $N_i=0$. Then (\ref{RNn})  become 
\be
R^{(n)}_k({\{0\}})&=&{ \gamma \,\left(cos\left( {2\pi\,k\over N} \right)-{g}\right)\over cos\left( {2\pi\,k\over N} \right)-{\gamma\, g}}\,,~~~~k=0,\cdot\cdot\cdot\cdot,\,N-1~~~~\label{ERn}
\ee
which reduces to formulas (\ref{ER0}) for N=2 case.  Likes as that in two-sites model  we expect that complexity will be divergent at critical value $g_c$, which indicates the quantum phase transition occurs at this point.  Let us discuss the property of  (\ref{ERn}). 
\\

1. As that in 2-sites system, above formula is  the exact value and we can let $n\to\infty$.  Being able to obtain the exact formula for ground state complexity is a special property in our method. 

2. We see that  $R^{(n\to\infty)}_{k=0}({\{0\}})$  in (\ref{ERn}) becomes zero at critical value $g=g_c=1$.  This implies that the associated complexity, $(D^{(n\to\infty)}_{(0,0)})^2$, is divergent and thus it is the quantum phase transition point. The critical value $g_c=1$ is same as that in two-sites system. 

3.  We plot the figure 4 to explicitly see this property.
\\
\\
\scalebox{.5}{\hspace{7cm}\includegraphics{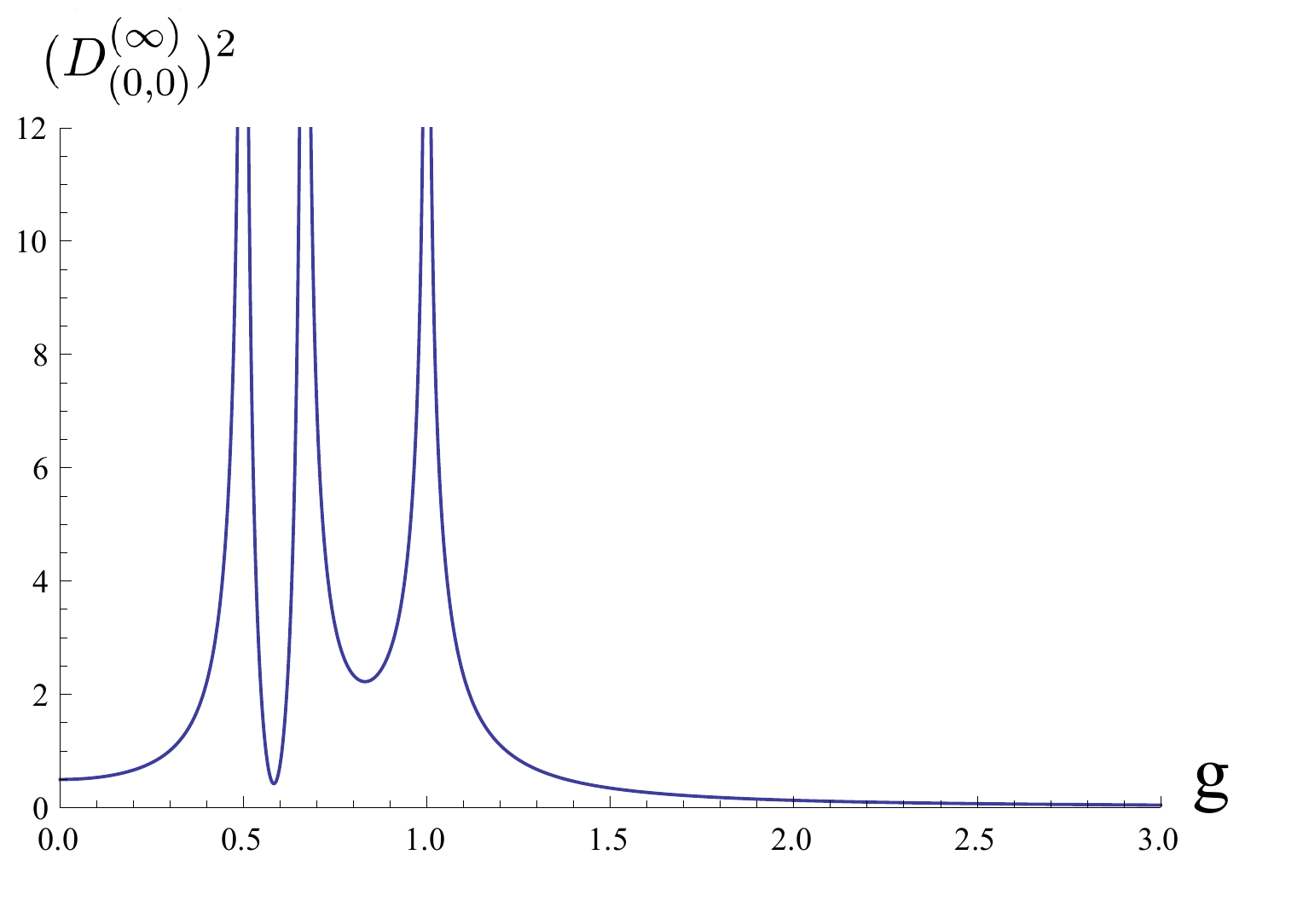}}
\\
{Figure 4:  Dependence of complexity $(D^{(n\to\infty)}_{(0,0)})^2$ on $``g"$ in 6-sites Bose-Hubbard model. The divergence at $g=g_c=1$ indicates the quantum phase transition at this point. The  figure is plotted with $\gamma_L=3/4$ while $\gamma_R=2$.}
\\

4. In two-sites system,  for range of $g>g_c$ we choose $\gamma=\gamma_R>1$ while for range of $g<g_c$ we choose $\gamma=\gamma_L<1$. In this algorithm, we see that  the complexity becomes infinity only at  $g=g_c$. In the N-sites system, however,  (\ref{ERn}) tells us that complexity could become infinity at the point $g=g_k<1$ if $cos\left( {2\pi\,k\over N} \right)-{g_k}=0$.  While we know that these are not physical critical points we have not yet found a simple  algorithm to rule it out. Figure 4 is plotted for 6-sites Bose-Hubbard model with $\gamma=2$.  The complexity for $g<1$ has  three divergence points which depend on $\gamma$ and are un-physical. 

5. When consider the excited state in which $\{N\}\ne\{0\}$ we can see that the associated complexity become divergent at the critical value which depends on the  reference state parameter $\gamma$. As that in 2-sites system the un-physical property reason may be that the model described in (\ref{bhmh}) is that keeps only the lowest vibrational state and could only describe the ground state property. 

6.  Generalize the method to N dimensional  Bose-Hubbard model is straightforward, as that in quantum field theory \cite{Jefferson1707}. We only need to  extend the N-sites closed chain Hamiltonian (\ref{bhmhN}) to 
\be
{\cal H}_N = {J\over 2} \sum_{\vec j} (\tilde a^\dag_{\vec j+\vec 1}\,\tilde a_{\vec j}+\tilde a^\dag_{{\vec j}-{\vec 1}}\tilde a_{\vec j})+ {U}\sum_{{\vec j}} \tilde n_{\vec j}\, (\tilde n_{\vec j} - 1),~~~~~~~\tilde n_{\vec j}=\tilde a^\dag_{\vec j}\,\tilde a_{\vec j}
\ee
in which ${\vec j}=(j_1,\,j_2,\,\cdot\cdot\cdot, j_N)$ is the N-dimensional vector with integral number. In this way, to find the formula of complexity we  only need to extend the single parameter Fourier transform to N parameters  transform and can get a general formula of complexity. For example the relation (\ref{FF}) becomes
\be
\(D^{(n)}_{\{N_i\}}\)^2=\sum_{\vec k}(N_{\vec k}+1)\,\left(\ln \(\sqrt {R^{(n)}_{\vec k}({\{N_i\}})}\)\right)^2,~~~~{\vec k}=(k_1,\,k_2,\,\cdot\cdot\cdot, k_N)
\ee
which is the complexity  of N dimensional  Bose-Hubbard model.


\section{Concluding Remarks}
We adopt  the operator approach to compute the complexity of the Bose-Hubbard Model.  Our prescription is to consider the value of $\<N_1\cdot\cdot\cdot N_n|{\cal H}|N_1\cdot\cdot\cdot N_n\>$. In this way, after proper mode expansion of  ${\cal H}$, many irrelevant terms become zero and the contributions of complexity from the remained terms could be calculated exactly through a perturbation algorithm. We  then obtain the  general formula of complexity therein.  It is interesting to see that the formula of complexity of Bose-Hubbard model is very similar to that of $\lambda\phi^4$ field theory.  Our analysis leads to a result that the critical point $g_c=1$ is found in 2-sites model and N-sites system.
The special property that the critical value being independent of lattice  number tells that our method is  mean-field like trick.  The more general method to compute complexity in many-body models remained to be found.

 Finally,  our algorithm can be applied to many quantum field theories and several many-body models in  condense matter. We will study the complexity of  Sachdev-Ye-Kitaev (SYK) model \cite{Sachdev2, Kitaev, Polchinski, Maldacena} in the next paper \cite{Huang2022} to see how the complexity probes the topological phase transition \cite{Fangli}.
\\
\begin{center} 
{\bf  \large References}
\end{center}

\end{document}